\documentclass[prl,groupedaddress,showpacs,twocolumn]{revtex4}
\usepackage[dvips]{graphicx}
\usepackage{amssymb}
\newcommand{\ped}[1]{\ensuremath{_{\rm #1}}}

\begin{document}

\title{The Magnetic-field Dependence of the Gaps in a Two-band Superconductor:
A Point-contact Study of MgB$_2$ Single Crystals}

\author{R.S. Gonnelli \email{E-mail:renato.gonnelli@polito.it},
D. Daghero, A. Calzolari, G.A. Ummarino, Valeria Dellarocca}

\affiliation{INFM - Dipartimento di Fisica, Politecnico di Torino,
Corso Duca degli Abruzzi 24, 10129 Torino, Italy}

\author{V.A. Stepanov}
\affiliation{P.N. Lebedev Physical Institute, Russian Academy of
Sciences, Leninski Pr. 53, 119991 Moscow, Russia}

\author{J. Jun, S.M. Kazakov and J. Karpinski}
\affiliation{Solid State Physics Laboratory, ETH, CH-8093
Z\"{u}rich, Switzerland}


\begin{abstract}
We present the results of directional point-contact measurements
in MgB$_2$ single crystals, in magnetic fields up to 9 T parallel
to the $c$ axis. By fitting the conductance curves of our point
contacts -- showing clear Andreev-reflection features -- with a
generalized Blonder-Tinkham-Klapwijk model, we were able to
extract the values of the two gaps $\Delta\ped{\sigma}$ and
$\Delta\ped{\pi}$. The comparison of the resulting
$\Delta\ped{\sigma}(B)$ and $\Delta\ped{\pi}(B)$ curves to the
theoretical predictions of the two-band model in dirty limit,
recently appeared in literature, allows the first direct test of
this model and gives a clear and quantitative proof that the $\pi$
band, even in the best single crystals, is in the moderate dirty
limit.
\end{abstract}
\maketitle

More than two years after the discovery of superconductivity in
MgB$_2$ \cite{Akimitsu} the great experimental and theoretical
efforts of the entire scientific community have led to
clarification of most features of this compound, mainly related to
the existence of two band systems ($\sigma$ and $\pi$) and of the
relevant gaps \cite{twoband,Brinkman}. Point-contact spectroscopy
(PCS) has proved particularly useful in investigating MgB$_2$,
since it allows measuring both the $\sigma$- and $\pi$-band gaps
at the same time \cite{szabo,nostroPRL} and determining the
temperature dependency of these gaps with very high accuracy
\cite{nostroPRL}. However, some controversial points are still
present. In particular, the effects of a magnetic field on the gap
values and on the $\pi$-band contribution to the total density of
states (DOS) are still under debate, as well as the clean or dirty
limit conditions of the two bands in the best samples and their
influence on the measured physical properties. Both these topics
can be addressed very well by performing accurate and directional
PCS experiments in high-quality MgB$_2$ single crystals in the
presence of magnetic field.

This paper presents some new results we obtained by PCS with
magnetic fields up to 9 T applied parallel to the \emph{c} axis of
MgB$_2$ single crystals. These measurements gave us the
magnetic-field dependence of the two gaps ($\Delta\ped{\pi}$ and
$\Delta\ped{\sigma}$) which results in good agreement with the
theoretical predictions for the mixed state in a dirty two-band
superconductor where the $\pi$-band diffusivity \emph{D}$_\pi$ is
a few times greater than the $\sigma$-band one, \emph{D}$_\sigma$
\cite{golubov}. The consistency of this result is tested by
comparing the zero-bias DOS calculated from our PCS data to the
averaged zero-bias DOS predicted by the same model \cite{golubov}.

The high-quality single crystals we used for our measurements were
produced at ETH (Zurich) by using the growth technique described
elsewhere~\cite{Karpinski}.  To control the direction of current
injection, we chose only samples with regular shape, i.e. flat
upper surface and sharp edges, even if these characteristics were
fulfilled only by smaller samples (up to about
0.5$\times$0.5$\times$0.07 mm$^3$). The point contacts were made
by using either a small spot of silver conductive paint or a small
piece of indium. The advantages of this technique with respect to
the standard one (that employs a metallic tip pressed against the
sample) have been widely explained elsewhere
\cite{nostroPRL,nostroPhysicaC}.
%Let us just point out here that,
%together with a much greater thermal stability of the contacts and
%reproducibility of their conductance curves, it allows a better
%control of the directionality of the current injection, since the
%surface is not damaged nor deformed. Because of their size
%($\varnothing \lesssim 50 \mu$m), our ``spot'' contacts should be
%rather considered as the parallel of several microscopic point
%contacts, and in fact their
The normal-state resistance of our contacts ranges from 10 to 50
$\Omega$. Due to the rather large mean free path of our single
crystals ($\ell \simeq 100$~nm) \cite{Sologubenko}, they result to
be in the ballistic regime \cite{nostroPhysicaC}.

In the experiments presented in this paper, the contacts were
always placed on the narrow side of the crystals, so that, as
already pointed out in previous papers
\cite{nostroPRL,nostroPhysicaC}, the probability of electron
injection was maximum along a direction parallel to the $ab$
planes. Thus, in the following we will refer to our contacts as
``$ab$-plane contacts''. The intensity of the magnetic field
applied to the samples could range from 0 up to 9 T and was
affected by an uncertainty of about 1\%, while the angle between
the field and the $ab$ planes was always $\phi=90\pm 2^{\circ}$.

The conductance curves ($\mathrm{d}I/\mathrm{d}V$ vs. $V$) of our
``spot'' contacts were obtained by numerical differentiation of
the $I-V$ characteristics, and then normalized to the normal-state
conductance.
\begin{figure}[t]
\vspace{-10mm}
\includegraphics[keepaspectratio, width=\columnwidth]{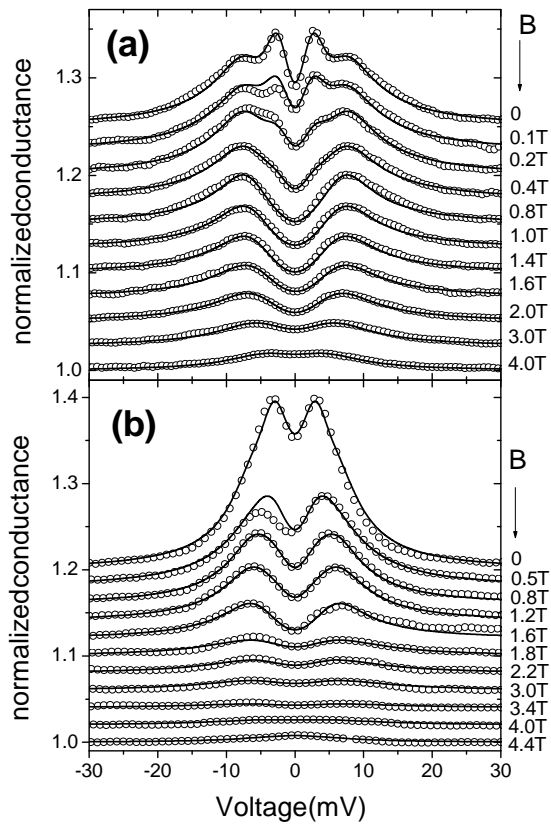}
\vspace{-10mm} \caption{(a) Normalized conductance curves of a
Ag-paint $ab$-plane contact, in increasing magnetic fields
parallel to the $c$ axis. Symbols: experimental data. Solid lines:
best-fitting curves given by the two-band BTK model (see text).
The curves are vertically shifted for clarity. (b) The same as in
(a) but for a In-spot $ab$-plane contact on a different
crystal.}\label{fig:1}
\end{figure}
The upper curve in Fig.\ref{fig:1}a (symbols) was obtained at
$T=4.2$~K in a Ag-paint contact. It has exactly the shape expected
for a $ab$-plane contact \cite{Brinkman}, featuring symmetric
conductance maxima at energies corresponding to
$\Delta\ped{\sigma}$ and $\Delta\ped{\pi}$. This curve can be
easily fitted by using a Blonder-Tinkham-Klapwijk (BTK) model
generalized to the two-band case, where the total
\emph{normalized} conductance is given by
$\sigma=(1-w\ped{\pi})\sigma\ped{\sigma}+w\ped{\pi}\sigma\ped{\pi}$
\cite{Brinkman,nostroPRL}. Here, both $\sigma\ped{\sigma}$ and
$\sigma\ped{\pi}$ are calculated in the case of a spherical Fermi
surface (FS), which is obviously not the case of MgB$_2$. This
problem cannot be overcome since, at present, there is no
analytical theory for the Andreev reflection that takes into
account the complex FS of MgB$_2$.

The fit has 7 free parameters: the gap amplitudes
$\Delta\ped{\pi}$ and $\Delta\ped{\sigma}$, the potential barrier
coefficients $Z\ped{\pi}$ and $Z\ped{\sigma}$, the lifetime
broadening parameters $\Gamma\ped{\pi}$ and $\Gamma\ped{\sigma}$
and, finally, the weight $w\ped{\pi}$. $\Gamma$ is a
phenomenological parameter that usually accounts for pair-breaking
effects (e.g. due to non-magnetic impurities), broadening the
conductance curves and reducing their height with respect to the
ideal ones. The best-fitting curve is shown in Fig.\ref{fig:1}a as
a solid line superimposed to the experimental data. The
corresponding values of the fitting parameters are:
$\Delta\ped{\sigma}=7.3\pm 0.1$ meV, $\Delta\ped{\pi}=2.8\pm 0.1$
meV, $Z\ped{\sigma}=0.944$, $Z\ped{\pi}=0.484$,
$\Gamma\ped{\sigma}=3.3$ meV, $\Gamma\ped{\pi}=1.46$ meV, and
finally $w\ped{\pi}=0.75$.  The gaps are in very good agreement
with the predictions of the two-band model (Eliashberg
formulation) \cite{Brinkman}. The same happens for the weight
$w\ped{\pi}$, that is compatible with an injection cone of about
26$^{\circ}$ \cite{nostroPRL}.

We can now try to fit in the same way the conductance curves
measured in the presence of magnetic field. Due to the lack of an
analytical theory of Andreev reflection for a two-band
superconductor in magnetic field, this fit can be simply performed
by using the lifetime broadening parameters
$\Gamma\ped{\sigma,\pi}$ to mimic the effect of the magnetic
field, as in Ref.~\onlinecite{Naidyuk}. In this case, the total
broadening parameter $\Gamma$ is considered as the sum of an
intrinsic $\Gamma\ped{i}$, due to the lifetime of quasiparticles,
and an extrinsic $\Gamma\ped{f}$ due to the effect of the magnetic
field: $\Gamma = \Gamma\ped{i} + \Gamma\ped{f}$. This approach
assumes that the field breaks pairs in a way that can be
represented by the quasiparticle lifetime, while its effects on
the DOS are negligible in a first-order approximation.

Unfortunately, at $T=0$~K the effects on the DOS curves of both
the magnetic field and the actual shape of the FS are very
relevant \cite{Dahmold,Dahmpriv,golubpriv}, so that our
BTK-lifetime model fails in fitting them. We expect similar large
effects to be also present in the Andreev-reflection conductance
at $T = 0$~K. Nevertheless, at low but finite temperature (for
example, $T \simeq 4$ K) the thermally-broadened theoretical DOS
curves become much more similar to those given by the standard
BTK-lifetime model. To clarify this point, we tried to fit with
this model some theoretical DOS curves at $T = 4$ K in the
presence of magnetic field parallel to the $c$ axis, i.e.: i)~the
local $\pi$-band DOS curves calculated at various points of the
vortex-lattice unit cell \cite{golubov}, ii)~the two-band DOS
calculated by taking into account the actual (simplified) shape of
the FS of MgB$_2$, made up of a distorted cylinder (for the
$\sigma$ band) and a half torus (for the $\pi$ band)
\cite{Dahm2,Dahmpriv}. In both cases the lifetime-broadened BTK
model proved very effective in fitting the theoretical curves in
all the energy range, apart from some deviations ($<30\%$) at very
small energies ($eV < 3$ meV). The gap values determined from the
fit are close to the theoretical ones used to generate the curves,
suggesting that $\Gamma\ped{f}$ can mimic rather well the
pair-breaking effect of the field at finite temperature.

\begin{figure}[t]
\begin{center}
\vspace{-2mm}
\includegraphics[keepaspectratio,width=0.8\columnwidth]{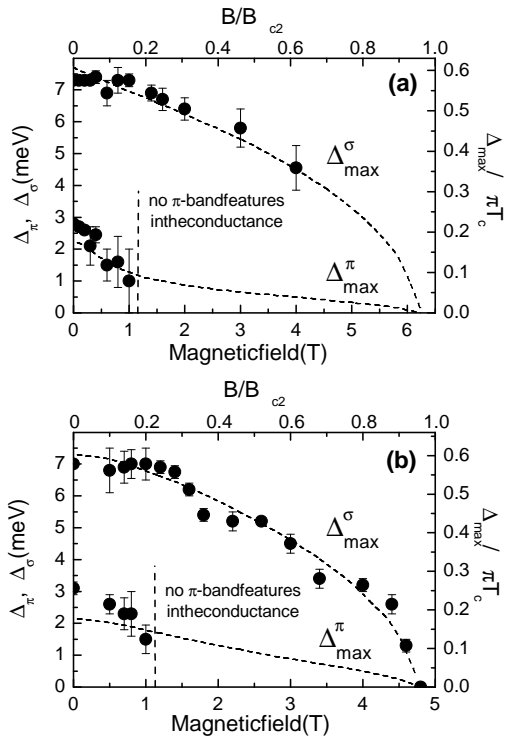}
\end{center}
\vspace{-6mm} \caption{(a) Symbols: magnetic field dependence of
the gaps determined from the curves of Fig.\ref{fig:1}a. The scale
is on the lower and left axes. Lines: maximum gap amplitudes
(calculated in Ref.\onlinecite{golubov} for $\mathbf{B}\parallel
c$) as a function of the normalized magnetic field, when
$D\ped{\sigma}=0.2 D\ped{\pi}$. The scale is on the upper and
right axes. (b) The same as in (a) but for the curves of
Fig.\ref{fig:1}b. The theoretical curves are calculated for
$D\ped{\sigma}=D\ped{\pi}$.} \label{fig:2}
\end{figure}

Thus, we can now proceed with the BTK fit of our experimental
curves at $B \neq 0$. Up to about 1~T the fit is carried out by
taking into account the contribution of the two bands, while above
this field the fit can be obtained with no contribution of the
$\pi$ band (i.e. by taking $\sigma\ped{\pi}=1$). The best value of
$w\ped{\pi}$ determined from the zero-field curve was kept
constant in all the other fits. As in most junctions, also the
values of $Z\ped{\sigma}$ and $Z\ped{\pi}$ could be taken
practically constant at the increase of the field, so that the
actual free fitting parameters at $B \neq 0$ are
$\Delta\ped{\sigma}$, $\Delta\ped{\pi}$, $\Gamma\ped{\sigma}$ and
$\Gamma\ped{\pi}$. The solid lines in Fig. \ref{fig:1}a represent
the curves that best fit the experimental data (circles). It is
clear that the fit is very good, even if it cannot reproduce the
asymmetry of some experimental curves.

In Fig.~\ref{fig:1}b the experimental conductances (circles) and
the corresponding best-fit curves (solid lines) are shown for
another junction obtained by pressing a very small In spot on the
side surface of a MgB$_2$ crystal. The best-fit parameters for the
zero-field curve are: $\Delta\ped{\sigma}=7.0\pm 0.1$ meV,
$\Delta\ped{\pi}=3.1\pm 0.2$ meV, $Z\ped{\sigma}=0.51$,
$Z\ped{\pi}=0.41$, $\Gamma\ped{\sigma}=3.0$ meV,
$\Gamma\ped{\pi}=1.86$ meV, and $w\ped{\pi}=0.7$. A comparison
with the corresponding values of Fig.\ref{fig:1}a shows that the
big difference in shape between the curves in (a) and (b) is
mainly due to the reduction of $Z\ped{\sigma}$ by a factor of two,
while all the other parameters are practically unchanged. This is
probably due to the different nature of the contact in the two
cases (Ag paint in (a), indium in (b)).

Figs.~\ref{fig:2}a and \ref{fig:2}b report the field-dependence of
the two gaps (solid circles) obtained by fitting the conductance
curves of Fig. \ref{fig:1}a and \ref{fig:1}b, respectively. It is
clearly seen that, for $B< 1$~T, the $\pi$-band gap decreases at
the increase of the field while the $\sigma$-band gap remains
practically unchanged. Above 1-1.2~T there is no longer trace of
features related to $\Delta\ped{\pi}$ in the conductance curves
and, therefore, \emph{nothing} can be concluded about this
gap~\footnote{In particular, we cannot say that
$\Delta\ped{\pi}=0$ for $B\geq B^*=1.2$~T. Thus, the
identification of $B^*$ (at which the $\pi$-band features
disappear) with $B\ped{c2}^{\pi}$ \cite{nostroPhysicaC} might not
be correct.}. The vanishing of the $\pi$-band contribution was
observed in \emph{all} the point contacts on MgB$_2$ crystals we
studied in magnetic field (more than 10). This result is in
complete agreement with many other experimental results present in
literature \cite{szabo,Eskildsen,Bouquet2} and with the
theoretical predictions of Ref. \onlinecite{golubov}, even if a
very recent paper claims the possibility to determine the
$\pi$-band gap up to about 5 T by PCS on MgB$_2$ films
\cite{Bugoslavsky}.

In both Figs.~\ref{fig:2}a and \ref{fig:2}b, dashed lines
represent the field dependence of the maximum pair potentials
($\Delta\ped{max}^{\sigma}$ and $\Delta\ped{max}^{\pi}$)
calculated at the boundary of the vortex-lattice unit cell for
$\mathbf{B}\parallel c$~\cite{golubov}. Even if our data rather
represent an average of the conductance in different regions of
the vortex lattice, the similarity between experimental data and
theoretical curves is remarkable, provided that the overall shape
of the curves is considered rather than the absolute gap values.
In Fig.\ref{fig:2}a, the initial decrease of $\Delta\ped{\pi}$ is
strikingly similar to that of $\Delta\ped{max}^{\pi}$, in the case
where $D\ped{\sigma}=0.2 D\ped{\pi}$. The behavior of
$\Delta\ped{\pi}$ in Fig.\ref{fig:2}b is more linear, and is here
compared to the theoretical curve calculated with
$D\ped{\sigma}=D\ped{\pi}$. As far as $\Delta\ped{\sigma}$ is
concerned, its field dependence agrees very well with the
theoretical curve in both cases. In Fig.\ref{fig:2}b, the
theoretical curve well reproduces the tendency of
$\Delta\ped{\sigma}$ to saturate at low fields.

The internal consistency of the results reported so far can be
checked by calculating the zero-bias DOS (ZBD) with the parameters
we got from the fits, and comparing its field dependence with that
reported in Ref.\onlinecite{golubov}. To do this, i) we took the
fitting parameters of each curve of Fig.~\ref{fig:1}; ii) we put
$Z\ped{\pi,\sigma}=20$ and assumed a negligible contribution of
the intrinsic lifetime broadening, i.e. we put
$\Gamma\ped{i}^{\pi, \sigma} =0$; iii) we calculated with the
lifetime-broadened BTK model the tunneling conductance curve at
$T=0$~K, by using the values of $\Delta\ped{\sigma}$,
$\Delta\ped{\pi}$, $w\ped{\pi}$ and $\Gamma\ped{f}^{\sigma, \pi}$
from the fit. The separation between intrinsic and field-induced
pair-breaking effects is possible since the former are reasonably
field-independent and can thus be determined from the fit of the
zero-field curves. This procedure is somehow similar to that used
in Ref.\onlinecite{Bouquet2} to evaluate the field-dependence of
$\gamma$, which is in fact a thermodynamic probe of the low-energy
DOS.
\begin{figure}[t]
\begin{center}
\includegraphics[width=0.7\columnwidth]{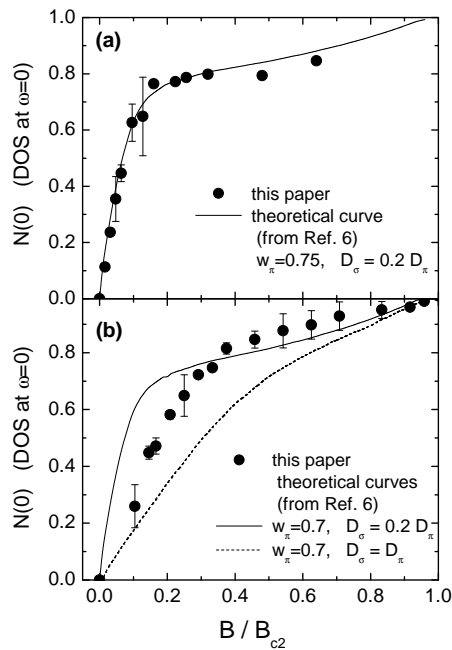}
\end{center}
\vspace{-6mm}\caption{(a) Symbols: magnetic field dependence of
the zero-bias DOS (ZBD) calculated with $\Delta\ped{\sigma,\pi}$
and $\Gamma\ped{f}^{\sigma,\pi}$ obtained by the fit of the curves
of Fig.\ref{fig:1}a, and taking $Z\ped{\sigma}=Z\ped{\pi}=20$.
Solid line: field dependence of the theoretical ZBD for
$w\ped{\pi}=0.75$, when $D\ped{\sigma}=0.2 D\ped{\pi}$,
\cite{golubov}. (b) Symbols: ZBD calculated as in (a) from the
curves of Fig.\ref{fig:1}b. Solid line: same as in (a) but with
$w\ped{\pi}=0.70$. Dashed line: field dependence of the ZBD when
$w\ped{\pi}=0.70$ and $D\ped{\sigma}=D\ped{\pi}$
\cite{golubov}.}\label{fig:3}
\end{figure}

In Figs.~\ref{fig:3}a and \ref{fig:3}b the values of the ZBD
determined from the data of Fig. \ref{fig:1}a and \ref{fig:1}b
(solid circles) are compared to the total theoretical averaged DOS
at $eV$=0 (solid and dashed curves) derived from the partial
$\pi$- and $\sigma$-band ZBD calculated in
Ref.\onlinecite{golubov}. In the case of Fig.~\ref{fig:3}a the
agreement between the data derived from the experiments and the
theoretical ZBD for $D\ped{\sigma} = 0.2 D\ped{\pi}$ is
impressive. On the other hand, the data of Fig.~\ref{fig:3}b lie
between the total ZBD for $D\ped{\pi} / D\ped{\sigma} = 5$ and
$D\ped{\pi} / D\ped{\sigma} = 1$, thus suggesting that an
intermediate value of this ratio would allow a proper fit of the
experimental data. Both these last results are fully consistent
with those shown in Fig.~\ref{fig:2}. However, this agreement is
not trivial since the quantities reported in Fig.~\ref{fig:2} and
Fig.~\ref{fig:3} are almost independent. In fact, it can be easily
shown that, for a given DOS curve, the value of the ZBD is little
dependent on the gap amplitude, but strongly depends on the values
of $\Gamma\ped{f}^{\pi}$ and $\Gamma\ped{f}^{\sigma}$. As a
consequence, the consistency of the two different comparisons
gives a strong evidence of the validity of our results.

To summarize, these results indicate that: i) the lifetime
broadening can be used, as a first approximation, to mimic the
effect of the magnetic field in the BTK fit of Andreev-reflection
curves measured in MgB$_2$ at finite temperature; ii) the field
dependencies of $\Delta\ped{\sigma,\pi}$ and of the ZBD deduced
from the experiments provide a clear and quantitative evidence
that, even in the best single crystals, the $\pi$ band is in
moderate dirty limit ($D\ped{\pi}\simeq 5 D\ped{\sigma}$), in
complete agreement with recent STM \cite{Eskildsen} and de
Haas-van Alphen \cite{Carrington} results on the same crystals, as
well as with the model for electric transport in
MgB$_2$~\cite{MazinPRL} and the theory of the mixed state in a
dirty two-band superconductor \cite{golubov}; iii) different
crystals, even if apparently produced in the same way, may have
different values of the diffusivity ratio $D\ped{\pi} /
D\ped{\sigma}$, i.e. different degrees of cleanliness of the $\pi$
band.

Some final comments are necessary on the zero-field values of
$\Delta\ped{\pi}$ and on the upper critical field
$B\ped{c2}^{\parallel c}$ implicitly reported in Fig.\ref{fig:2}.
First, our PCS values of $\Delta\ped{\pi}$ ($\simeq 2.7-2.8$
meV)\cite{nostroPRL} are systematically greater than those given
by STM ($\simeq 2.1-2.2$ meV) \cite{Eskildsen}. Since the accuracy
of both the techniques is out of discussion, this difference has
probably a physical reason that is still under investigation. Let
us just remind here that the values of $\Delta\ped{\pi}$ are
distributed between 1 and 3 meV over the FS~\cite{Choi}, with a
maximum at about 2 meV. Moreover, strong-coupling
calculations~\cite{Brinkman} give an average value over the FS
$\Delta\ped{\pi} = 2.7$ meV. This might suggest that, while STM is
more sensitive to the maximum of the distribution, PCS gives a
weighed average of $\Delta\ped{\pi}$ over the FS. As far as the
critical field $B\ped{c2}$ is concerned, we recently determined it
by PCS for both $\mathbf{B}\parallel c$ and $\mathbf{B} \perp
c$~\cite{nostroRIO}. We always got values greater than those of
$B\ped{c2}$ measured by bulk techniques (torque magnetometry,
specific heat) and smaller than the surface critical field
$B\ped{c3}$~\cite{nostroRIO}. The results will be widely discussed
in a forthcoming paper, but we can anticipate that the high values
of $B\ped{c2}$ measured by PCS are probably related to some
non-trivial surface effects.

In conclusion, in this paper we have presented the first detailed
determination of the magnetic-field dependency of the gaps in a
two-band superconductor by using the point-contact technique in
high-quality MgB$_2$ single crystals. The results allow the first
direct test of the recent theory for the mixed state in a dirty
two-band superconductor. Moreover, they give clear and
self-consistent evidence of the moderate dirty-limit conditions
for the $\pi$ band and of the variability of these conditions from
sample to sample.

We are indebted to T. Dahm, O.V. Dolgov, A.A. Golubov and I.I.
Mazin for discussions and suggestions. V.A.S. acknowledges the
support from RFBR (project N. 02-02-17133), the Ministry of
Science and Technologies of the Russian Federation (contract N.
40.012.1.1.1357) and the INTAS project N.01-0617. This work was
done within the project PRA "UMBRA" of INFM and the ASI contract
N. I/R/109/02.

\end{document}